\documentclass[twocolumn,aps,pra,amsmath,amssymb]{revtex4-1}
\usepackage[latin9]{inputenc}
\usepackage{mathrsfs}
\usepackage{amstext}
\usepackage{amssymb}
\usepackage{graphicx}
\usepackage{esint}

\makeatletter
\usepackage{epsfig}
\usepackage{bm}
\usepackage{dcolumn}
\usepackage{color}

\makeatother

\begin{document}

\title{Quantum fluctuations of a resonantly interacting $p$-wave Fermi
superfluid in two dimensions}

\author{Hui Hu$^{1}$, Brendan C. Mulkerin$^{1}$, Lianyi He$^{2}$, Jia
Wang$^{1}$, and Xia-Ji Liu$^{1}$ }

\affiliation{$^{1}$Centre for Quantum and Optical Science, Swinburne University
of Technology, Melbourne, Victoria 3122, Australia}

\affiliation{$^{2}$Department of Physics and State Key Laboratory of Low-Dimensional
Quantum Physics, Tsinghua University, Beijing 100084, China}

\date{\today}
\begin{abstract}
Using the Gaussian pair fluctuation theory, we investigate quantum
fluctuations of a strongly interacting two-dimensional chiral \textit{p}-wave
Fermi superfluid at the transition from a Bose-Einstein condensate
(BEC) to a topologically non-trivial Bardeen-Cooper-Schrieffer (BCS)
superfluid. Near the topological phase transition at zero chemical
potential, $\mu=0$, we observe that quantum fluctuations strongly
renormalize the zero-temperature equations of state, sound velocity,
pair-breaking velocity, and Berezinskii-Kosterlitz-Thouless (BKT)
critical temperature of the Fermi superfluid, all of which can be
non-analytic functions of the interaction strength. The indication
of non-analyticity is particularly evident in the BKT critical temperature,
which also exhibits a pronounced peak near the topological phase transition.
Across the transition and towards the BEC limit we find that the system
quickly becomes a trivial interacting Bose liquid, whose properties
are less dependent on the interparticle interaction. The qualitative
behavior of composite bosons in the BEC limit remains to be understood. 
\end{abstract}

\maketitle

\section{Introduction}

Unconventional electronic superconductivity and fermionic superfluidity
are of great interest and lie at the heart of many intriguing quantum
materials \cite{Mineev1999}. One of the most important examples is
the two-dimensional (2D) chiral $p$-wave superconductor (superfluid),
where the pairing order parameter has the $p_{x}+ip_{y}$ symmetry
in its orbital angular momentum. It was shown to be topologically
non-trivial with vortex excitations that exhibit non-Abelian statistics
\cite{Read2000,Ivanov2001}. These so-called Majorana excitations
have been suggested to be a key ingredient for processing topological
quantum computation \cite{Kitaev2003,Nayak2008}. Unfortunately, in
spite of extensive search for decades, a 2D $p$-wave superconductor
remains elusive in condensed matter physics. The best-known candidate
material of 2D $p$-wave superconductors so far is strontium ruthenate
Sr$_{2}$RuO$_{4}$, whose superconductivity was first observed by
Maeno and his group in 1994 \cite{Maeno1994}.

The recent realization of resonantly interacting ultracold atomic
Fermi gases opens a new paradigm to create the topological $p$-wave
superfluid \cite{Gurarie2007}. By tuning the $s$-wave interparticle
interaction in a two-component Fermi gas through magnetic Feshbach
resonances, the crossover from a Bardeen-Cooper-Schrieffer (BCS) fermionic
superfluid to a Bose-Einstein condensate (BEC) has now been routinely
observed in laboratories \cite{Bloch2008,Giorgini2008}, confirming
the long-sought BEC-BCS crossover \cite{Eagles1969,Leggett1980,Nozieres1985,SadeMelo1993}
in both three and two dimensions. A resonantly interacting $p$-wave
Fermi gas can be realized by either using $p$-wave Feshbach resonances
or by preparing fermionic atoms in the same hyperfine pseudo-spin
state, which experience long-range dipole-dipole interactions. The
former has already been demonstrated for $^{40}$K and $^{6}$Li atoms
\cite{Regal2003,Zhang2004,Gunter2005,Schunck2005,Gaebler2007,Fuchs2008,Inada2008,Nakasuji2013,Luciuk2016,Waseem2017,Yoshida2018,Wassem2018},
although the system suffers a serious loss in atom number near the
$p$-wave resonance. Nevertheless, in three dimensions the system
can still reach a quasi-equilibrium state \cite{Luciuk2016}, in which
a number of interesting physical properties of the cloud can be experimentally
examined. More importantly, in lower dimensions the atom loss has
been found to be significantly reduced \cite{Waseem2017}, as theoretically
predicted \cite{Levinsen2008,Fedorov2017}. For 
 a single-component dipolar Fermi gas \cite{Lu2012,Aikawa2014} the
$s$-wave scattering is completely suppressed by Pauli exclusion principle.
The $p$-wave component of the interparticle interaction could then
be significantly enhanced by suitably tuning the strength of the dipole-dipole
interaction. All these recent experimental advances in ultracold atoms
make the realization of a 2D $p$-wave Fermi superfluid a very appealing
idea.

Theoretically, the many-body physics of strongly interacting $p$-wave
Fermi gases has been studied to some extent \cite{Gurarie2007}. These
include the exploration of the phase diagram \cite{Botelho2005,Ho2005,Gurarie2005,Cheng2005,Iskin2006,Cao2013},
which becomes richer due to the anisotropy in the different $p$-wave
channels, determining the transition temperature for the superfluid
transition in three dimensions \cite{Ohashi2005,Inotani2012,Inotani2015}
or the Berezinskii-Kosterlitz-Thouless (BKT) transition in two dimensions
\cite{Cao2017}, as well as the calculation of the $p$-wave contact
parameters \cite{Yoshida2015,Yu2015,He2016,Peng2016,Zhang2017,Yao2018,Inotani2018},
which characterize the universal short-distance and large-momentum
behavior of the system \cite{Tan2008,Braaten2008}. Most of these
theoretical investigations rely on the mean-field theory, which qualitatively
captures the underlying physics of the $p$-wave pairing. To describe
more accurately a $p$-wave Fermi superfluid, in particular in two
dimensions, it is necessary to include strong quantum fluctuations
beyond mean-field close to the resonantly interacting regime \cite{Liu2018,Jiang2018}.
In this respect, it is convenient to adopt the Gaussian pair fluctuation
(GPF) theory \cite{Hu2006,Diener2008}, which provides a quantitatively
reliable description of an $s$-wave Fermi superfluid at the BEC-BCS
crossover, in both three \cite{Hu2006,Hu2007,Diener2008} and two
dimensions \cite{He2015}.

In this work, we explore quantum fluctuations in a 2D chiral $p$-wave
Fermi superfluid using the GPF theory, paying specific attention to
the role played by the topological phase transition at zero chemical
potential. A number of physical observables at zero temperature are
considered across the BEC-BCS transition, such as the chemical potential,
total energy, pressure equation of state, sound velocity, pair-breaking
velocity, and also the critical velocity for superfluidity. All these
quantities are strongly affected by quantum fluctuations. By assuming
the existence of well-defined fermionic Bogoliubov quasi-particles
and bosonic excitations of phonons, we further calculate the temperature
dependence of superfluid fraction with the approximate Landau formalism
\cite{Bighin2016}. This leads to an improved determination of the
BKT critical temperature in the strongly interacting regime.

The paper is organized as follows. In the next section (Sec. II),
we present the model Hamiltonian of a 2D spin-less $p$-wave interacting
Fermi gas. In Sec. III, we describe the details of the GPF theory
of the chiral $p$-wave Fermi superfluid. In Sec. IV, we first discuss
various equations of state as a function of the interaction strength,
including the chemical potential, pressure, and total energy. We then
present the results of sound velocity, pair-breaking velocity, and
critical velocity. Based on the single-particle fermionic excitation
spectrum and the sound velocity at zero temperature, we calculate
the temperature dependence of superfluid density within the Landau
picture for superfluidity and consequently determine the BKT critical
temperature. Finally, in Sec. VI we give our conclusions and outlook.

\section{Model Hamiltonian}

We consider a spin-less 2D atomic Fermi gas of density $n$, interacting
in the dominant $p$-wave channel near a broad $p$-wave Feshbach
resonance, as described by a single-channel Hamiltonian (we set the
area $A=1$) \cite{Botelho2005}, 
\begin{equation}
{\cal H}=\sum_{{\bf k}}\xi_{{\bf k}}\psi_{{\bf k}}^{\dagger}\psi_{{\bf k}}+\frac{1}{2}\sum_{{\bf k},{\bf k}^{\prime},{\bf q}}V_{\mathbf{k}\mathbf{k}'}b_{\mathbf{k}\mathbf{q}}^{\dagger}b_{\mathbf{k'}\mathbf{q}},
\end{equation}
where $\psi_{{\bf k}}$ ($\psi_{{\bf k}}^{\dagger}$) is the annihilation
(creation) field operator for atoms with mass $M$ and the single-particle
dispersion $\xi_{{\bf k}}\equiv\epsilon_{{\bf k}}-\mu=\hbar^{2}\mathbf{k}^{2}/(2M)-\mu$,
and $b_{\mathbf{k}\mathbf{q}}\equiv\psi_{-{\bf k}{\bf +q}/2}\psi_{{\bf k}{\bf +q}/2}$
is the composite operator that annihilates a pair of atoms with a
center-of-mass momentum $\mathbf{q}$. We work with the grand-canonical
ensemble and tune the chemical potential $\mu$ to make the average
density 
\begin{equation}
\sum_{\mathbf{k}}\left\langle \hat{n}_{{\bf k}}\right\rangle =n\equiv\frac{k_{F}^{2}}{4\pi},
\end{equation}
where $\hat{n}_{\mathbf{k}}\equiv\psi_{{\bf k}}^{\dagger}\psi_{{\bf k}}$
and $k_{F}$ is the Fermi wave-vector. For the inter-particle interaction,
we adopt the following separable form \cite{Nozieres1985,Botelho2005,Ho2005},
\begin{equation}
V_{\mathbf{k}\mathbf{k}'}=\lambda\Gamma\left({\bf k}\right)\Gamma^{*}\left({\bf k'}\right),
\end{equation}
where $\lambda<0$ is the \emph{bare} interaction strength and the
dimensionless regularization function $\Gamma\left({\bf k}\right)$
represents the chiral $p_{x}+ip_{y}$ symmetry of the pairing interaction,
i.e.,

\begin{equation}
\Gamma\left({\bf k}\right)=\frac{\left(k/k_{F}\right)}{\left[1+\left(k/k_{0}\right)^{2n}\right]^{3/2}}e^{i\varphi_{{\bf k}}}.
\end{equation}
Here, $k_{0}$ is a large momentum cut-off, which is necessary to
make the model Hamiltonian renormalizable, and $\varphi_{{\bf k}}$
is the polar angle of ${\bf k}$. We use the exponent $n$ to tune
the shape of the regularization function $\Gamma\left({\bf k}\right)$
and to confirm the insensitivity of our results on the form of the
interparticle interaction. The choice of $n=1/2$ was used earlier
by Noziéres and Schmitt-Rink \cite{Nozieres1985}, and Botelho and
Sá de Melo \cite{Botelho2005}. In this paper, unless otherwise specified,
we follow the work by Ho and Diener \cite{Ho2005} and take $n=1$
for the numerical results presented. Actually, the results depend
very weakly on the exponent $n$. The use of other values of $n$
only leads to small quantitative difference.

In principle, the bare interaction strength $\lambda$ and the cut-off
momentum $k_{0}$ should be renormalized (i.e., replaced) in terms
of the 2D $p$-wave scattering area $a_{p}$ and effective range $R_{p}\sim1/k_{0}$
\cite{Zhang2017}. However, for a better presentation, it turns out
to be more convenient to use a scattering energy $E_{b}$ \cite{Botelho2005,Cao2017},
which is basically the ground state energy of two fermions at zero
center-of-mass momentum, 
\begin{equation}
2\epsilon_{{\bf k}}\psi_{{\bf k}}+\sum_{{\bf k}^{\prime}}V_{{\bf kk}^{\prime}}\psi_{{\bf k}^{\prime}}=E_{b}\psi_{{\bf k}}.
\end{equation}
By inserting the separable interaction potential, it is easy to obtain,
\begin{equation}
\frac{1}{\lambda}=-\sum_{{\bf k}}\frac{\left|\Gamma\left({\bf k}\right)\right|^{2}}{2\epsilon_{{\bf k}}-E_{b}}.\label{eq:renorm}
\end{equation}
We note that, unlike the $s$-wave scattering in 2D, where the scattering
energy $E_{b}$ is always negative, in our $p$-wave case $E_{b}$
can be either negative or positive. A negative scattering energy indicates
the existence of a two-body bound state (i.e., on the BEC side), with
a binding energy $\varepsilon_{B}=-E_{b}>0$. On the other hand, the
weakly interacting BCS limit is reached at $E_{b}\rightarrow+\infty$.
Throughout the paper, we use the set of parameters ($E_{b},k_{0},n=1$)
to characterize the $p$-wave interaction. Their relation to the $p$-wave
scattering area $a_{p}$ and effective range $R_{p}$ is briefly discussed
in Appendix A.

\section{Gaussian pair function theory at zero temperature}

In the superfluid phase at zero temperature, it is useful to introduce
the Nambu spinor presentation for the field operators \cite{Hu2006,Diener2008},
\begin{equation}
\Psi_{{\bf k}}=\left(\begin{array}{l}
\psi_{{\bf k}}\\
\psi_{-{\bf k}}^{\dagger}
\end{array}\right),
\end{equation}
with which, the model Hamiltonian can be rewritten as, 
\begin{equation}
{\cal H}=\frac{1}{2}\sum_{{\bf k}}\Psi_{{\bf k}}^{\dagger}\left(\xi_{{\bf k}}\sigma_{z}\right)\Psi_{{\bf k}}+\frac{1}{2\lambda}\sum_{{\bf q}}\hat{\rho}_{\mathbf{q}}^{\dagger}\hat{\rho}_{\mathbf{q}},
\end{equation}
where 
\begin{equation}
\hat{\rho}_{\mathbf{q}}\equiv\lambda\sum_{{\bf k}}\Gamma^{*}\left({\bf k}\right)b_{\mathbf{kq}}=\lambda\sum_{{\bf k}}\Psi_{{\bf k-\frac{q}{2}}}^{\dagger}\Gamma^{*}\left({\bf k}\right)\sigma_{-}\Psi_{{\bf k+\frac{q}{2}}}
\end{equation}
is a generalized density operator for a pair of fermions and, $\sigma_{z}$
and $\sigma_{\pm}=(\sigma_{x}\pm\sigma_{y})/2$ are the Pauli matrices.
In the following, we first solve the model Hamiltonian at the mean-field
level and then include Gaussian pair fluctuations on top of the mean-field
solution.

\subsection{Mean-field theory}

The superfluid phase is characterized by a nonzero (real) pairing
order parameter $\Delta$ at zero center-of-mass momentum $\mathbf{q}=0$,
i.e., 
\begin{equation}
\hat{\rho}_{\mathbf{q}}=\Delta\delta_{\mathbf{q},\mathbf{0}}+\Delta_{\mathbf{q}},
\end{equation}
where $\Delta_{\mathbf{q}}$ is the pair fluctuation field around
the order parameter. Inserting this decoupling into the model Hamiltonian,
we obtain 
\begin{eqnarray}
\mathscr{\mathcal{H}} & = & \mathcal{H}_{\textrm{MF}}+\frac{1}{2\lambda}\sum_{{\bf q\neq0}}\Delta_{\mathbf{q}}^{\dagger}\Delta_{\mathbf{q}},\\
\mathcal{H}_{\textrm{MF}} & = & \frac{1}{2}\sum_{{\bf k}}\Psi_{{\bf k}}^{\dagger}\left[\begin{array}{cc}
\xi_{{\bf k}} & \Delta\Gamma\left({\bf k}\right)\\
\Delta\Gamma^{*}\left({\bf k}\right) & -\xi_{{\bf k}}
\end{array}\right]\Psi_{{\bf k}}-\frac{\Delta^{2}}{2\lambda}.\label{eq:hamiMF}
\end{eqnarray}
Here, we neglect the fluctuation field at zero momentum, which gives
a vanishing contribution in the thermodynamic limit. The mean-field
Hamiltonian can be straightforwardly solved by diagonalizing the two
by two matrix in Eq. (\ref{eq:hamiMF}). This leads to the following
energy of Bogoliubov quasi-particles, 
\begin{equation}
E_{{\bf k}}=\sqrt{\xi_{{\bf k}}^{2}+\Delta^{2}\left|\Gamma\left({\bf k}\right)\right|^{2}},
\end{equation}
and quasi-particle wave-functions, 
\begin{eqnarray}
\left|u_{{\bf k}}\right|^{2} & = & \frac{1}{2}\left(1+\frac{\xi_{{\bf k}}}{E_{{\bf k}}}\right),\\
\left|v_{{\bf k}}\right|^{2} & = & \frac{1}{2}\left(1-\frac{\xi_{{\bf k}}}{E_{{\bf k}}}\right),\\
u_{{\bf k}}v_{{\bf k}}^{*} & = & \frac{\Delta\Gamma\left({\bf k}\right)}{2E_{{\bf k}}}.
\end{eqnarray}
The BCS Green function 
\begin{equation}
{\cal G}_{0}\left({\bf k},i\omega_{m}\right)=\left[\begin{array}{ll}
i\omega_{m}-\xi_{{\bf k}} & -\Delta\Gamma\left({\bf k}\right)\\
-\Delta\Gamma^{*}\left({\bf k}\right) & i\omega_{m}+\xi_{{\bf k}}
\end{array}\right]^{-1},
\end{equation}
where $\omega_{m}=(2m+1)\pi k_{B}T$ ($m\in\mathbb{Z}$) is the fermionic
Matsubara frequency, is then given by,

\begin{eqnarray}
{\cal G}_{0}^{11}\left({\bf k},i\omega_{m}\right) & = & \frac{u_{{\bf k}}u_{{\bf k}}^{*}}{i\omega_{m}-E_{{\bf k}}}+\frac{v_{{\bf k}}v_{{\bf k}}^{*}}{i\omega_{m}+E_{{\bf k}}},\\
{\cal G}_{0}^{12}\left({\bf k},i\omega_{m}\right) & = & \frac{u_{{\bf k}}v_{{\bf k}}^{*}}{i\omega_{m}-E_{{\bf k}}}-\frac{u_{{\bf k}}v_{{\bf k}}^{*}}{i\omega_{m}+E_{{\bf k}}},\\
{\cal G}_{0}^{21}\left({\bf k},i\omega_{m}\right) & = & \frac{u_{{\bf k}}^{*}v_{{\bf k}}}{i\omega_{m}-E_{{\bf k}}}-\frac{u_{{\bf k}}^{*}v_{{\bf k}}}{i\omega_{m}+E_{{\bf k}}},\\
{\cal G}_{0}^{22}\left({\bf k},i\omega_{m}\right) & = & \frac{v_{{\bf k}}v_{{\bf k}}^{*}}{i\omega_{m}-E_{{\bf k}}}+\frac{u_{{\bf k}}u_{{\bf k}}^{*}}{i\omega_{m}+E_{{\bf k}}}.
\end{eqnarray}
The pairing order parameter can be determined by minimizing the mean-field
thermodynamic potential, 
\begin{eqnarray}
\Omega_{\text{MF}} & = & \frac{1}{2}\frac{\Delta^{2}}{\lambda}+\frac{1}{2}\sum_{{\bf k}}\left(\xi_{{\bf k}}-E_{{\bf k}}\right),\nonumber \\
 & = & \frac{1}{2}\sum_{{\bf k}}\left[\xi_{{\bf k}}-E_{{\bf k}}-\frac{\Delta^{2}\left|\Gamma\left({\bf k}\right)\right|^{2}}{2\epsilon_{{\bf k}}-E_{b}}\right].\label{eq:OmegaMF}
\end{eqnarray}
Thus, we obtain the gap equation, 
\begin{equation}
\sum_{{\bf k}}\left[\frac{1}{2E_{{\bf k}}}-\frac{1}{2\epsilon_{{\bf k}}-E_{b}}\right]\left|\Gamma\left({\bf k}\right)\right|^{2}=0.\label{eq:gapMF}
\end{equation}
At the mean-field level, as mentioned earlier, the chemical potential
$\mu$ is adjusted to satisfy the mean-field number equation, 
\begin{equation}
n=n_{F}\equiv-\frac{\partial\Omega_{\text{MF}}}{\partial\mu}=\frac{1}{2}\sum_{{\bf k}}\left(1-\frac{\xi_{{\bf k}}}{E_{{\bf k}}}\right).\label{eq:numMF}
\end{equation}

\begin{figure}
\centering{}\includegraphics[width=0.43\textwidth]{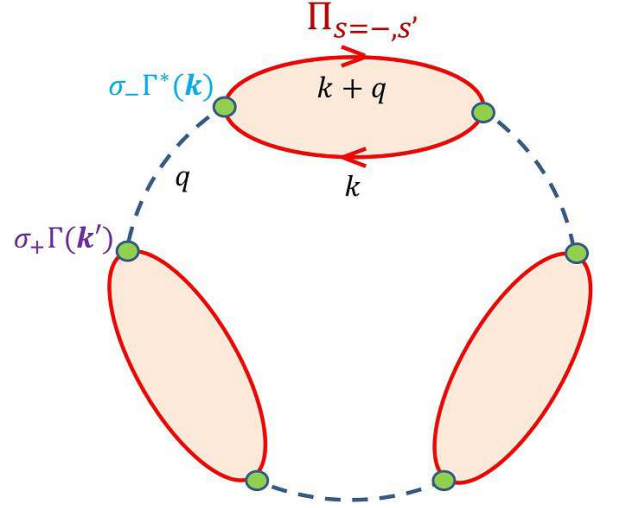}\caption{\label{fig1_diagram} (color online). The third order ladder diagram
considered in the Gaussian pair fluctuation theory. The solid line
with arrow represents the two by two BCS Green functions. The dashed
line with two vertices indicates the interparticle interaction. There
are four types of ladders, $\Pi_{ss'}$, depending on the choice of
the interaction vertex: $s=+$ for $\sigma_{+}\Gamma(\mathbf{k}')$
and $s=-$ for $\sigma_{-}\Gamma^{*}(\mathbf{k})$.}
\end{figure}

\subsection{Gaussian pair fluctuation theory}

We now take into account the fluctuation terms $\Delta_{\mathbf{q}}^{\dagger}\Delta_{\mathbf{q}}/(2\lambda)$
at nonzero center-of-mass momentum. At the lowest Gaussian level,
their contribution to the thermodynamic potential can be represented
by the ladder (or bubble) diagrams \cite{Nozieres1985,Hu2006}, one
of which (i.e., the third order diagram) is shown in Fig. \ref{fig1_diagram},
where the dashed lines denote the bare interaction $\lambda\Gamma({\bf k)}\Gamma^{*}({\bf k'})$.
Following the standard diagrammatic rules \cite{Abrikosov1963}, a
$n$-th order ladder diagram gives the following contribution to the
thermodynamic potential, 
\begin{eqnarray}
\Omega_{\text{GF}}^{(n)} & = & \frac{\left(-1\right)^{n+1}\lambda^{n}}{2n}\sum_{\mathcal{Q}}\sum_{s_{1},\cdots,s_{n}'}\left[\Pi\left(\mathcal{Q}\right)\right]_{s_{1}s_{1}'}\cdots\left[\Pi\left(\mathcal{Q}\right)\right]_{s_{n}s_{n}'},\nonumber \\
 & = & \frac{\left(-1\right)^{n+1}\lambda^{n}}{2n}\sum_{\mathcal{Q}}\textrm{Tr}\left[\begin{array}{cc}
\Pi_{-+}\left(\mathcal{Q}\right) & \Pi_{--}\left(\mathcal{Q}\right)\\
\Pi_{++}\left(\mathcal{Q}\right) & \Pi_{+-}\left(\mathcal{Q}\right)
\end{array}\right]^{n}\label{eq:OmegaGFn}
\end{eqnarray}
where we have used the short-hand notations $\mathcal{Q}=(\mathbf{q},i\nu_{n})$
with $\nu_{n}=2n\pi k_{B}T$ ($n\in\mathbb{Z}$) being the bosonic
Matsubara frequency, and $\sum_{\mathcal{Q}}\equiv k_{B}T\sum_{i\nu_{n}}\sum_{\mathbf{q}}$.
The subscript $s=-,+$ (or $s'$) of the pair propagator $\left[\Pi\left(\mathcal{Q}\right)\right]_{s,s'}$
stands for the interaction vertex $\sigma_{-}\Gamma^{*}(\mathbf{k})$
and $\sigma_{+}\Gamma(\mathbf{k'})$, respectively. The different
choice for $s$ and $s'$ leads to four kinds of ladders and hence
four pair propagators:
\begin{widetext}
\begin{eqnarray}
\Pi_{-+}\left(\mathcal{Q}\right) & = & k_{B}T\sum_{{\bf k},i\omega_{m}}\left|\Gamma\left({\bf k}\right)\right|^{2}\text{Tr}\left[\sigma_{-}{\cal G}_{0}\left({\bf k}+\frac{{\bf q}}{2},i\omega_{m}\right)\sigma_{+}{\cal G}_{0}\left({\bf k}-\frac{{\bf q}}{2},i\omega_{m}-i\nu_{n}\right)\right],\\
\Pi_{--}\left(\mathcal{Q}\right) & = & k_{B}T\sum_{{\bf k},i\omega_{m}}\left[\Gamma^{*}\left({\bf k}\right)\right]^{2}\text{Tr}\left[\sigma_{-}{\cal G}_{0}\left({\bf k}+\frac{{\bf q}}{2},i\omega_{m}\right)\sigma_{-}{\cal G}_{0}\left({\bf k}-\frac{{\bf q}}{2},i\omega_{m}-i\nu_{n}\right)\right],
\end{eqnarray}
$\Pi_{++}\left(\mathcal{Q}\right)=\left[\Pi_{--}\left(\mathcal{Q}\right)\right]^{*}$,
and $\Pi_{+-}\left(\mathcal{Q}\right)=\left[\Pi_{-+}\left(\mathcal{Q}\right)\right]^{*}$.
However, the summation indices $s_{1},\cdots,s_{n}'$ in $\Omega_{\text{GF}}^{(n)}$
can not take arbitrary values. As each interaction line contains the
vertex $\sigma_{-}\Gamma^{*}(\mathbf{k})$ and $\sigma_{+}\Gamma(\mathbf{k'})$
in \emph{pairs}, we must have $s_{i}'=-s_{i+1}$ for $i=1,\cdots,n$
(we set $n+1\rightarrow1$). The summation over the vertex indices
therefore leads to the trace of a matrix product, as given in Eq.
(\ref{eq:OmegaGFn}). The contribution of all the ladder diagrams
is then readily to calculate, by summing over $n$. We find that,

\begin{equation}
\Omega_{\text{GF}}\left[\mu,\Delta\left(\mu\right)\right]=\frac{1}{2}\sum_{\mathcal{Q}}\textrm{Tr}\ln\left[-\frac{1}{\lambda}+\left(\begin{array}{ll}
\Pi_{-+}\left(\mathcal{Q}\right) & \Pi_{--}\left(\mathcal{Q}\right)\\
\Pi_{++}\left(\mathcal{Q}\right) & \Pi_{+-}\left(\mathcal{Q}\right)
\end{array}\right)\right]\equiv\frac{1}{2}\sum_{\mathcal{Q}}\ln\det\left[\begin{array}{ll}
M_{11}\left(\mathcal{Q}\right) & M_{12}\left(\mathcal{Q}\right)\\
M_{21}\left(\mathcal{Q}\right) & M_{22}\left(\mathcal{Q}\right)
\end{array}\right]\label{eq:OmegaGF}
\end{equation}
where the explicit expression of $\mathbf{M}(\mathcal{Q})$ is given
by, 
\begin{eqnarray}
M_{11}\left(\mathcal{Q}\right) & = & \sum_{{\bf k}}\left|\Gamma_{\mathbf{k}}\left({\bf k}\right)\right|^{2}\left[\frac{\left(u_{+}u_{+}^{*}\right)\left(u_{-}u_{-}^{*}\right)}{i\nu_{n}-E_{+}-E_{-}}-\frac{\left(v_{+}v_{+}^{*}\right)\left(v_{-}v_{-}^{*}\right)}{i\nu_{n}+E_{+}+E_{-}}+\frac{1}{2E_{\mathbf{k}}}\right],\\
M_{12}\left(\mathcal{Q}\right) & = & \sum_{{\bf k}}\left[\Gamma^{*}\left({\bf k}\right)\right]^{2}\left[\frac{\left(u_{+}v_{+}^{*}\right)\left(u_{-}v_{-}^{*}\right)}{i\nu_{n}-E_{+}-E_{-}}-\frac{\left(u_{+}v_{+}^{*}\right)\left(u_{-}v_{-}^{*}\right)}{i\nu_{n}+E_{+}+E_{-}}\right],
\end{eqnarray}
\end{widetext}

$M_{21}(\mathcal{Q})=M_{12}^{*}(\mathcal{Q})$, and $M_{22}(\mathcal{Q})=M_{11}^{*}(\mathcal{Q})$.
Here, we abbreviate $u_{\pm}\equiv u_{{\bf q}/2\pm{\bf k}}$, $v_{\pm}\equiv v_{{\bf q}/2\pm{\bf k}}$,
and $E_{\pm}\equiv E_{{\bf q}/2\pm{\bf k}}$, and rewrite the bare
interaction strength $\lambda$ using Eq. (\ref{eq:renorm}) and Eq.
(\ref{eq:gapMF}). In $\Omega_{\textrm{GF}}$, we have exchanged the
order of the trace and ``ln'' operators, which gives rise to the
determinant of the pair propagator matrix. Moreover, the summation
over the bosonic Matsubara frequency $i\nu_{n}$ diverges, as a result
of $M_{11}(\mathcal{Q})\sim\nu_{n}^{1/2}$ in the limit of $\nu_{n}\rightarrow\infty$.
This divergence can be formally cured by imposing a convergence factor
and converting the summation into a contour integral along the real
axis \cite{Hu2006}. In practice, it is more convenient to adopt an
interesting trick proposed by Diener and his co-workers at zero temperature
\cite{Diener2008}. We define the regular part of the pair propagators
$M_{11}(\mathcal{Q})$ and $M_{22}(\mathcal{Q})$ \cite{Diener2008,He2015}:
\begin{equation}
M_{11}^{C}=\sum_{{\bf k}}\left|\Gamma\left({\bf k}\right)\right|^{2}\left[\frac{\left(u_{+}u_{+}^{*}\right)\left(u_{-}u_{-}^{*}\right)}{i\nu_{n}-E_{+}-E_{-}}+\frac{1}{2E_{\mathbf{k}}}\right],
\end{equation}
and $M_{22}^{C}(\mathcal{Q})=[M_{11}^{C}(\mathcal{Q})]^{*}$. It is
easy to check that $M_{11}^{C}(\mathbf{q},i\nu_{n}\rightarrow z)$
has no singularities or zeros (i.e., poles and branch cuts) in the
left-half complex plane of $\textrm{Re}\,z<0$, as a result of $\left|u_{\pm}\right|^{2}\leq1$
and $E_{+}+E_{-}\geq2E_{\mathbf{k}}$. At zero temperature, we obtain
\begin{equation}
k_{B}T\sum_{i\nu_{n}}\ln M_{11}^{C}\left(\mathcal{Q}\right)=k_{B}T\sum_{i\nu_{n}}\ln M_{22}^{C}\left(\mathcal{Q}\right)=0,
\end{equation}
after writing them in terms of a standard contour integral \cite{Diener2008}.
Therefore, we arrive at \cite{Diener2008} 
\begin{equation}
\Omega_{\text{GF}}=\frac{1}{2}\sum_{\mathbf{q}}k_{B}T\sum_{i\nu_{n}}\ln\frac{\left[M_{11}M_{22}-M_{12}M_{21}\right]\left(\mathcal{Q}\right)}{M_{11}^{C}\left(\mathcal{Q}\right)M_{22}^{C}\left(\mathcal{Q}\right)}.
\end{equation}
A further simplification can be made by noticing that, at zero temperature
($T\rightarrow0$), we may take $\nu_{n}\rightarrow\omega$ as a continuous
variable and rewrite the summation $k_{B}T\sum_{i\nu_{n}}$ in the
form of an integral, $\int_{-\infty}^{+\infty}d\omega/(2\pi)$ \cite{Diener2008,He2015}.
By defining the following five functions \cite{He2015},

\begin{eqnarray}
M_{11}^{C} & = & A\left({\bf q},\omega\right)-i\omega B\left({\bf q},\omega\right),\\
M_{11}-M_{11}^{C} & = & -\Delta^{4}C\left({\bf q},\omega\right)+i\omega\Delta^{4}D\left({\bf q},\omega\right),\\
M_{12} & = & 2\Delta^{2}F\left({\bf q},\omega\right),
\end{eqnarray}
the Gaussian fluctuation contribution to the thermodynamic potential
finally takes the form,
\begin{widetext}
\begin{equation}
\Omega_{\text{GF}}\left[\mu,\Delta\left(\mu\right)\right]=\int\limits _{0}^{\infty}\frac{d\omega}{2\pi}\sum_{{\bf q}}\ln\left[1-2\Delta^{4}\left(\mu\right)\frac{AC+\omega^{2}BD+2F^{2}}{A^{2}+\omega^{2}B^{2}}+\Delta^{8}\left(\mu\right)\frac{C^{2}+\omega^{2}D^{2}}{A^{2}+\omega^{2}B^{2}}\right].
\end{equation}
The explicit form of the five functions is given by, 
\begin{eqnarray}
A\left({\bf q},\omega\right) & = & -\frac{1}{4}\sum_{{\bf k}}\left|\Gamma\left({\bf k}\right)\right|^{2}\left[\left(\frac{1}{E_{+}}+\frac{1}{E_{-}}\right)\frac{\left(E_{+}+\xi_{+}\right)\left(E_{-}+\xi_{-}\right)}{\omega^{2}+\left(E_{+}+E_{-}\right)^{2}}-\frac{2}{E}\right],\\
B\left({\bf q},\omega\right) & = & +\frac{1}{4}\sum_{{\bf k}}\left|\Gamma\left({\bf k}\right)\right|^{2}\frac{1}{E_{+}E_{-}}\frac{\left(E_{+}+\xi_{+}\right)\left(E_{-}+\xi_{-}\right)}{\omega^{2}+\left(E_{+}+E_{-}\right)^{2}},\\
C\left({\bf q},\omega\right) & = & +\frac{1}{4}\sum_{{\bf k}}\left|\Gamma\left({\bf k}\right)\right|^{2}\left|\Gamma\left(\frac{{\bf q}}{2}+{\bf k}\right)\right|^{2}\left|\Gamma\left(\frac{{\bf q}}{2}-{\bf k}\right)\right|^{2}\left(\frac{1}{E_{+}}+\frac{1}{E_{-}}\right)\frac{1}{\left(E_{+}+\xi_{+}\right)\left(E_{-}+\xi_{-}\right)}\frac{1}{\omega^{2}+\left(E_{+}+E_{-}\right)^{2}},\\
D\left({\bf q},\omega\right) & = & +\frac{1}{4}\sum_{{\bf k}}\left|\Gamma\left({\bf k}\right)\right|^{2}\left|\Gamma\left(\frac{{\bf q}}{2}+{\bf k}\right)\right|^{2}\left|\Gamma\left(\frac{{\bf q}}{2}-{\bf k}\right)\right|^{2}\frac{1}{E_{+}E_{-}}\frac{1}{\left(E_{+}+\xi_{+}\right)\left(E_{-}+\xi_{-}\right)}\frac{1}{\omega^{2}+\left(E_{+}+E_{-}\right)^{2}},\\
F\left({\bf q},\omega\right) & = & -\frac{1}{4}\sum_{{\bf k}}\left[\Gamma^{*}\left({\bf k}\right)\right]^{2}\Gamma\left(\frac{{\bf q}}{2}+{\bf k}\right)\Gamma\left(\frac{{\bf q}}{2}-{\bf k}\right)\left(\frac{1}{E_{+}}+\frac{1}{E_{-}}\right)\frac{1}{\omega^{2}+\left(E_{+}+E_{-}\right)^{2}}.
\end{eqnarray}
\end{widetext}

For our case with a chiral $p$-wave interaction (i.e., $\Gamma(k)\propto k_{x}+ik_{y}$),
one may show that the above five functions do not depend on the polar
angle of ${\bf q}$ (see Appendix B), and thus we can simply set ${\bf q}=q{\bf e}_{x}$
in the $\mathbf{k}$-integration of $A$, $B$, $C$, $D$ and $F$.
This reduces the calculation of $\Omega_{\text{GF}}$ to a four-dimensional
integration (over $\omega$, $q=\left|\mathbf{q}\right|$, $k=\left|\mathbf{k}\right|$
and $\varphi_{\mathbf{k}}$).

For a given chemical potential $\mu$, once the fluctuation thermodynamic
potential $\Omega_{\textrm{GF}}$ is obtained, we calculate the number
of Cooper pairs $n_{B}$ by using numerical differentiation, 
\begin{equation}
2n_{B}=-\frac{\partial\Omega_{\textrm{GF}}\left[\mu,\Delta\left(\mu\right)\right]}{\partial\mu}.
\end{equation}
Within the GPF theory, we then adjust the chemical potential to satisfy
the number equation $n=n_{F}+2n_{B}$. It is worth noting that the
pairing gap $\Delta\left(\mu\right)$ is always determined at the
mean-field level by using the gap equation, Eq. (\ref{eq:gapMF}),
in order to have a gapless Goldstone phonon mode \cite{Hu2006,Diener2008}.

\section{Results and discussions}

For the convenience of numerical calculations we take the Fermi wave-vector
$k_{F}$ as the units of the wave-vectors ($k,q$), and the Fermi
energy $\varepsilon_{F}=\hbar^{2}k_{F}^{2}/(2m)=2\pi\hbar^{2}n/m$
as the units of energy and temperature. This is equivalent to setting
$2m=\hbar=k_{B}=1$. In the following, we mainly choose a cut-off
momentum $k_{0}=30k_{F}$ and the dependence of various properties
on $k_{0}$ is briefly discussed at the end of the section.

\subsection{Equation of state}

\begin{figure}
\centering{}\includegraphics[width=0.48\textwidth]{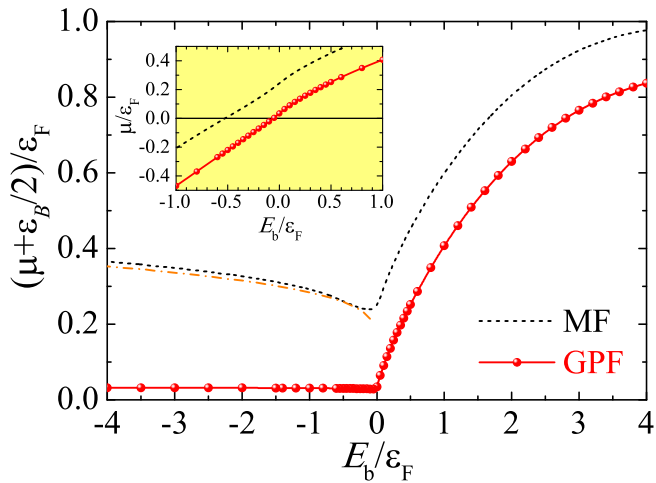}\caption{\label{fig2_mu} (color online). The chemical potential $\mu$ as
a function of the scattering energy $E_{b}$, calculated by using
the mean-field theory (dashed line) and GPF theory (solid line with
circles). The dot-dashed line shows the asymptotic behavior in the
BEC limit predicted by the mean-field theory in Eq. (\ref{eq:gBMF}),
which corresponds to a constant molecular scattering length for composite
bosons. In the main figure, we have subtracted the contribution from
the two-body bound state with the binding energy $\varepsilon_{B}\equiv\max(-E_{b},0)$.
The inset highlights the chemical potential near the topological phase
transition (i.e., $\mu\sim0$ or $E_{b}\sim0$). Here, we take a cut-off
momentum $k_{0}=30k_{F}$.}
\end{figure}

In Fig. \ref{fig2_mu}, we report the chemical potential $\mu$ as
a function of the interaction strength $E_{b}$, predicted by the
mean-field theory and GPF theory. To clearly show the many-body effect,
we have subtracted the two-body contribution from the bound state
when the scattering energy $E_{b}<0$, which takes the form $-\varepsilon_{B}/2\equiv-\max(-E_{b},0)/2$.
In the BCS limit ($E_{b}\gg\varepsilon_{F}$), both mean-field and
GPF theories predict $\mu\rightarrow\varepsilon_{F}$, as expected.
However, towards the BEC limit ($E_{b}\ll-\varepsilon_{F}$), they
show entirely different behavior.

In the BEC limit we anticipate that the system may turn into a weakly
interacting Bose condensate of composite Cooper pairs, with a bosonic
chemical potential given by, 
\begin{equation}
\mu_{B}=2\mu+\varepsilon_{B}\simeq g_{B}n_{B},
\end{equation}
where $n_{B}\simeq n/2$ and $g_{B}$ is the strength of the interaction
between two Cooper pairs. Physically, $g_{B}$ should decrease as
we move to the BEC limit. Using the relation $\varepsilon_{F}=2\pi\hbar^{2}n/m$,
we obtain that 
\begin{equation}
g_{B}\simeq\left(\frac{8\pi\hbar^{2}}{m}\right)\frac{\mu+\varepsilon_{B}/2}{\varepsilon_{F}}.
\end{equation}
Thus, we observe from Fig. \ref{fig2_mu} that the mean-field theory
incorrectly predicts an increasing pair-pair interaction strength
when we approach the BEC limit, while the GPF theory gives a \emph{small}
residual pair-pair interaction, which is essentially independent on
the scattering energy $E_{b}$.

In the mean-field theory, the pair-pair interaction strength can be
analytically calculated using a Ginzburg-Landau free energy functional
for the pair fluctuation field $\Delta_{\mathbf{q}}$ (see Appendix
C). We find that, 
\begin{equation}
g_{B,\textrm{MF}}=\frac{16\pi\hbar^{2}}{m}\frac{\left[\ln\eta+2\eta^{-1}-\eta^{-2}/2-3/2\right]}{\left(\ln\eta+\eta^{-1}-1\right)^{2}},\label{eq:gBMF}
\end{equation}
where $\eta=\hbar^{2}k_{0}^{2}/(m\left|E_{b}\right|)+1$. As $\eta\gg1$
for the parameters in Fig. \ref{fig2_mu}, to a good approximation
we have 
\begin{equation}
g_{B,\textrm{MF}}\simeq\frac{16\pi\hbar^{2}}{m}\frac{1}{\ln\left[\hbar^{2}k_{0}^{2}/\left(m\left|E_{b}\right|\right)\right]},
\end{equation}
which explains the wrong behavior of stronger pair-pair interaction
as we decrease $E_{b}$ (see the dot-dashed line in Fig.~\ref{fig2_mu}).

Quite generally, the mean-field theory breaks down in two dimensions
due to enhanced quantum fluctuations. This is already known for an
$s$-wave Fermi superfluid \cite{He2015}, where the mean-field theory
predicts a \emph{constant} pair-pair interaction strength of $4\pi\hbar^{2}/m$,
instead of a much smaller and chemical potential dependent coupling
strength. The renormalization of the pair-pair interaction due to
quantum fluctuations is well-captured by our GPF theory. Indeed, in
an $s$-wave Fermi superfluid the GPF theory is reliable in predicting
an accurate molecular scattering length for composite bosons \cite{He2015},
in good agreement with the exact four-body calculation and diffusion
quantum Monte Carlo (QMC) simulation. In our case of a chiral $p$-wave
Fermi superfluid, we anticipate that the GPF theory will similarly
lead to a reliable result for the pair-pair interaction strength $g_{B}$.
Unfortunately, unlike the $s$-wave Fermi superfluid, the existence
the regularization function $\Gamma\left({\bf k}\right)$ makes it
infeasible to derive an analytic expression for $g_{B}$. In future
studies, the QMC calculation of the ground-state energy of the system
or the exact solution of four resonantly $p$-wave interacting fermions
in two dimensions would be very useful to understand the small and
constant pair-pair interaction strength $g_{B}$, as predicted by
our GPF theory.

\begin{figure}
\centering{}\includegraphics[width=0.48\textwidth]{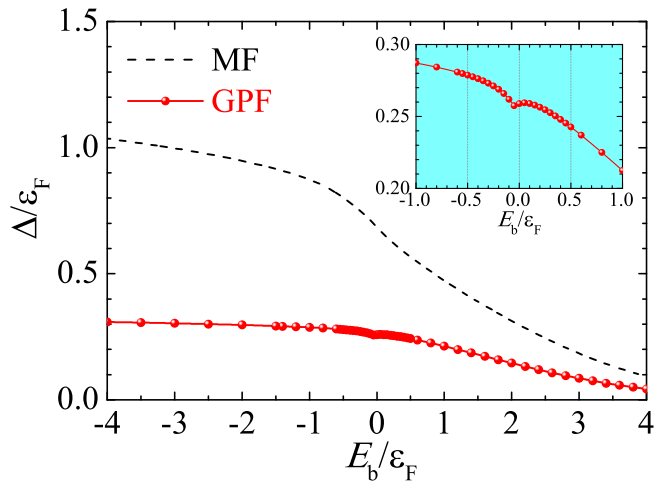}\caption{\label{fig3_gap} (color online). The pairing gap $\Delta$ as a function
of the scattering energy $E_{b}$, predicted by using the mean-field
theory (dashed line) and GPF theory (solid line with circles). The
inset highlights the kink in the pairing gap near the topological
phase transition (i.e., $E_{b}\sim0$). Here, we take a cut-off momentum
$k_{0}=30k_{F}$.}
\end{figure}

\begin{figure}
\centering{}\includegraphics[width=0.48\textwidth]{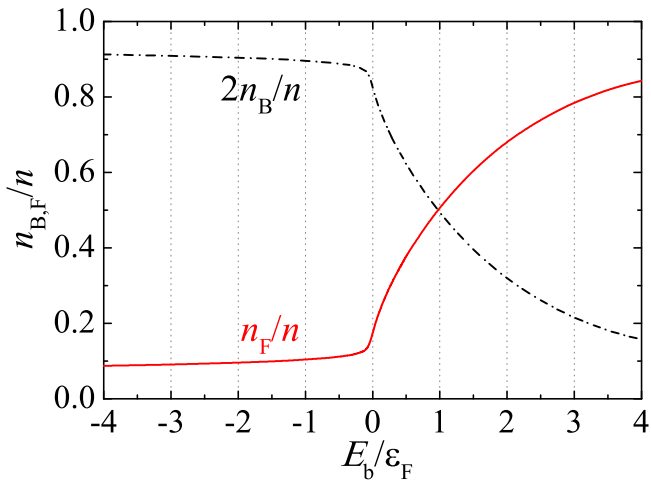}\caption{\label{fig4_num} (color online). The fraction of fermions $n_{F}/n$
(solid line) and Cooper pairs $n_{B}/n$ (dashed line), as a function
of the scattering energy $E_{b}$. Here, we take a cut-off momentum
$k_{0}=30k_{F}$.}
\end{figure}

Let us now consider the intermediate coupling regime near zero scattering
energy $E_{b}=0$, where the chemical potential $\mu$ changes sign
and the system is expected to undergo a topological phase transition.
In sharp contrast to the $s$-wave case, where $\mu$ evolves rather
smoothly, here we find a dramatic change in the slope of the quantity
$\mu+\varepsilon_{B}/2$ at $E_{b}\sim0$ or $\mu\sim0$. This non-analytic
feature at the topological phase transition has been noticed in previous
mean-field studies \cite{Botelho2005,Cao2013,Cao2017} and we see
that quantum fluctuations make it even more pronounced.

Figure \ref{fig3_gap} presents the evolution of the pairing order
parameter $\Delta$ as a function of the scattering energy $E_{b}$,
calculated using the mean-field theory (dashed line) and the GPF theory
(solid line with circles). Away from the BCS limit, the pairing gap
is significantly reduced by quantum fluctuations. In particular, at
resonance, the pairing gap is about a quarter of the Fermi energy,
$\Delta\sim0.25\varepsilon_{F}$. There is an apparent dip at the
topological phase transition, as a result of the non-analyticity of
the thermodynamics at the transition.

Theoretically, the significance of quantum fluctuations can be most
easily recognized from the evolution of the number of Cooper pairs
$n_{B}$ as a function of the scattering energy $E_{b}$, as shown
in Fig. \ref{fig4_num}. We find a rapid increase in $n_{B}$, when
we move to the topological phase transition point from the BCS limit.
Upon reaching the transition, the dependence of the number of Cooper
pairs on the scattering energy becomes nearly flat. Once again, this
may be viewed as an indication of the non-analyticity at the topological
phase transition.

\begin{figure}
\centering{}\includegraphics[width=0.48\textwidth]{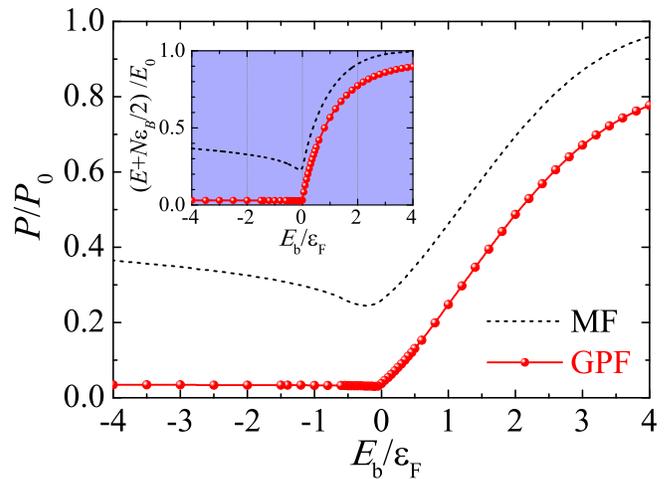}\caption{\label{fig5_pressure} (color online). The pressure (main figure)
and total energy (inset), as a function of the scattering energy $E_{b}$,
predicted by using the mean-field theory (dashed line) and GPF theory
(solid line with circles). The pressure and energy are normalized
with respect to the ideal gas values $P_{0}=n\varepsilon_{F}/2$ and
$E_{0}=N\varepsilon_{F}/2$, respectively. Here, $N$ is the total
number of particles. For the total energy, we have subtracted the
contribution from the two-body bound state, $-N\varepsilon_{B}/2$.
As before, we take a cut-off momentum $k_{0}=30k_{F}$.}
\end{figure}

In experiments, on the other hand, the non-analyticity of the thermodynamic
functions at the transition may be probed by measuring the \emph{homogeneous}
pressure equation of state through the density distribution of a harmonically
trapped resonant $p$-wave Fermi superfluid \cite{Nascimbene2010}.
In Fig. \ref{fig5_pressure}, we report the pressure $P$, normalized
to its non-interacting value $P_{0}=n\varepsilon_{F}/2$, as a function
of the scattering energy $E_{b}$, calculated with the mean-field
theory and the GPF theory. The pressure $P$ shows almost the same
scattering energy dependence as the chemical potential, with a clear
kink at the topological phase transition. Therefore, the observation
of this kink may be regarded as an \emph{indirect} proof the topological
phase transition \cite{TopoSuperfluidNote}. Moreover, the measurement
of the small and nearly constant pressure on the BEC side will be
useful to clarify the nature of the resulting weak-interacting Bose
condensate.

To conclude this subsection, it is worth noting a recent study of
the same system by Jiang and Zhou \cite{Jiang2018}, based on a two-channel
model for a broad $p$-wave resonance. In that study, quantum fluctuations
from selected two-loop diagrams are found to destabilize the system
at the resonance, in disagreement with our finding of a stable Fermi
superfluid at all interaction strengths. This discrepancy is unlikely
from the different model Hamiltonian (i.e., one-channel vs. two-channel),
since the one-channel model and two-channel model are known to give
the same description for a broad Feshbach resonance \cite{Diener2004,Liu2005}.
It should come from the treatment of quantum fluctuations at different
levels. The GPF treatment presented in this work, when it is generalized
to the two-channel model \cite{Ohashi2003}, includes the two-loop
diagrams selected by Jiang and Zhou \cite{Jiang2018,TwoLoopsDiagramNote}.
Moreover, it may pick up a set of marginal diagrams containing higher-order
loops, within the ladder or bubble approximation. A future GPF study
of the two-channel model for a resonantly interacting $p$-wave Fermi
superfluid will be useful to clarify the discrepancy and to provide
more accurate results for a narrow Feshbach resonance.

\subsection{Critical velocity for superfluidity}

A superfluid loses its superfluidity when it moves faster than a critical
velocity. For an $s$-wave Fermi superfluid, the critical velocity
in the BCS and BEC limits is given by the pair-breaking velocity and
sound velocity, respectively, and exhibits a maximum in between \cite{Diener2008}.
A maximum critical velocity at the resonance emphasizes the stability
of a strongly interacting Fermi superfluid \cite{Toniolo2017}.

\begin{figure}
\centering{}\includegraphics[width=0.48\textwidth]{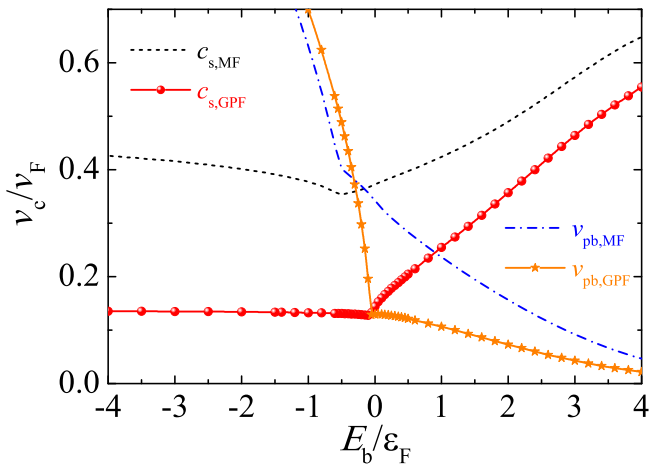}\caption{\label{fig6_vc} (color online). The sound velocity and pair-breaking
velocity, as a function of the scattering energy $E_{b}$, predicted
by using the mean-field theory (dashed line or dash-dotted line) and
GPF theory (solid line with circles or stars). As the scattering energy
decreases (or the interaction strength increases), the critical velocity
$v_{c}=\max\{v_{pb},c_{s}\}$ slowly increases. Here, we take a cut-off
momentum $k_{0}=30k_{F}$.}
\end{figure}

The situation for a $p$-wave Fermi superfluid seems to be a bit different.
In Fig. \ref{fig6_vc}, we present the sound velocity determined from
the equation of state, 
\begin{equation}
c_{s}=\left[\frac{n}{m}\frac{\partial\mu}{\partial n}\right]^{1/2}=\left[\frac{n}{m}\left(-\frac{\partial^{2}\Omega}{\partial\mu^{2}}\right)^{-1}\right]^{1/2},
\end{equation}
and the pair-breaking velocity calculated by using Landau criterion,
\begin{equation}
v_{pb}=\min_{\{\mathbf{k}\}}\frac{E_{\mathbf{k}}}{\left|k\right|}\simeq\left\{ \begin{array}{cc}
\Delta & \textrm{if }\mu\geq0\\
\sqrt{\Delta^{2}-4\mu} & \textrm{if }\mu<0
\end{array}\right..
\end{equation}
In both mean-field and GPF frameworks, the resulting critical velocity
$v_{c}=\max\{v_{pb},c_{s}\}$ roughly increases with decreasing scattering
energy $E_{b}$. In particular, on the BEC side, the GPF result of
the critical velocity becomes nearly flat, consistent with a constant
pair-pair interaction strength observed earlier. Typically, the critical
velocity at resonance is about $0.1v_{F}$, smaller than that of an
$s$-wave Fermi superfluid \cite{Bighin2016,Mulkerin2017}. This means
that a $p$-wave Fermi superfluid could be more easily destroyed than
its $s$-wave counterpart.

\subsection{BKT transition temperature}

In two dimensions, the transition to a superfluid state at finite
temperature is governed by the BKT mechanism \cite{Berezinskii1972,Kosterlitz1973}.
The BKT critical temperature $T_{c}$ of a chiral $p$-wave Fermi
superfluid was considered in the previous studies by using the mean-field
theory \cite{Cao2017}. Here, we determine $T_{c}$ with the inclusion
of quantum fluctuations.

\begin{figure}
\centering{}\includegraphics[width=0.48\textwidth]{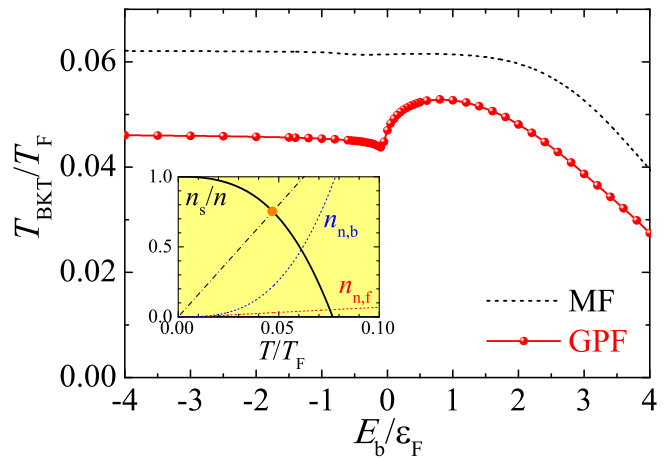}\caption{\label{fig7_tc} (color online). The BKT transition temperature, calculated
by using the mean-field theory (dashed line) and GPF theory (solid
line with circles), as a function of the scattering energy $E_{b}$.
The inset shows the normal density fractions contributed by fermions
$n_{n,f}/n$ and by pairs $n_{n,b}/n$, and the superfluid density
fraction $n_{s}/n=1-n_{n,f}/n-n_{n,b}/n$ at zero scattering energy
$E_{b}=0$. The cross point between the superfluid fraction $n_{s}/n$
and the line $16T/T_{F}$ determines the BKT temperature at $E_{b}=0$.
Here, we take a cut-off momentum $k_{0}=30k_{F}$.}
\end{figure}

For this purpose, we need to calculate the superfluid density $n_{s}$
and then determine $T_{c}$ using the so-called Thouless-Nelson criterion
\cite{Nelson1977}, 
\begin{equation}
k_{B}T_{c}=\frac{\pi\hbar^{2}}{8m}n_{s}\left(T_{c}\right),
\end{equation}
or equivalently, 
\begin{equation}
\frac{T_{c}}{T_{F}}=\frac{1}{16}\frac{n_{s}\left(T_{c}\right)}{n}.
\end{equation}
A full calculation of superfluid density $n_{s}$ within the GPF framework
is numerically involved. Here, we follow the idea by Bighin and Salasnich
to approximately calculate the superfluid density using the standard
Landau formalism \cite{Bighin2016}. This provides an approximate
but convenient way to include quantum fluctuations \cite{Bighin2016,Mulkerin2017}.

To apply the Landau formalism, we assume that the low-energy excitations
of the resonantly interacting $p$-wave superfluid are well-described
by quasi-particles. This assumption is excellent in both BCS and BEC
limits. Therefore, we anticipate that it may also give some qualitative
predictions near resonance. Following Landau's quasi-particle picture
\cite{Khalatnikov2000}, the densities of the normal fluid, due to
single-particle fermionic excitations and collective bosonic excitations,
are respectively given by, 
\begin{eqnarray}
n_{n,f} & = & -\frac{\hbar^{2}}{m}\sum_{{\bf k}}\frac{{\bf k}^{2}}{2}\frac{\partial}{\partial E_{{\bf k}}}\left(\frac{1}{e^{E_{{\bf k}}/k_{B}T}+1}\right),\\
n_{n,b} & = & -\frac{\hbar^{2}}{m}\sum_{{\bf q}}\frac{{\bf q}^{2}}{2}\frac{\partial}{\partial\omega_{{\bf q}}}\left(\frac{1}{e^{\omega_{{\bf q}}/k_{B}T}-1}\right),
\end{eqnarray}
where we approximate that, as a rough estimation, the fermionic excitations
have the energy spectrum of $E_{{\bf k}}$ and the bosonic excitations
have phonon dispersion $\omega_{{\bf q}}\simeq c_{s}q$. The superfluid
density $n_{s}$ then takes the form, 
\begin{equation}
n_{s}=n-n_{n,f}-n_{n,b}.
\end{equation}

At resonance, the normal densities due to fermionic and bosonic excitations,
$n_{n,f}$ and $n_{n,b}$, and the superfluid density $n_{s}$ are
shown in the inset of Fig. \ref{fig7_tc}. We find that the bosonic
degree of freedom gives the dominant contribution to the superfluid
density and hence leads to a reduced BKT critical temperature. Indeed,
the mean-field theory predicts a nearly saturated critical temperature
$T_{c}=T_{F}/16\sim0.06T_{F}$ at resonance, while our GPF theory
with Landau formalism for superfluid density gives a smaller critical
temperature $T_{c}\sim0.04T_{F}$.

In the main figure of Fig. \ref{fig7_tc}, we present the evolution
of the BKT critical temperature $T_{c}$ as a function of the scattering
energy $E_{b}$. It exhibits a bump near the resonance with a maximum
$T_{c,\max}\simeq0.052T_{F}$ at $E_{b}\sim\varepsilon_{F}$. The
cusp at $E_{b}\simeq0$ may be viewed as a clear demonstration of
the non-analyticity of the finite temperature thermodynamics at the
topological phase transition. Towards the BEC limit, we find that
the BKT critical temperature saturates to $T_{c}\sim0.047T_{F}$.

\begin{figure}
\centering{}\includegraphics[width=0.48\textwidth]{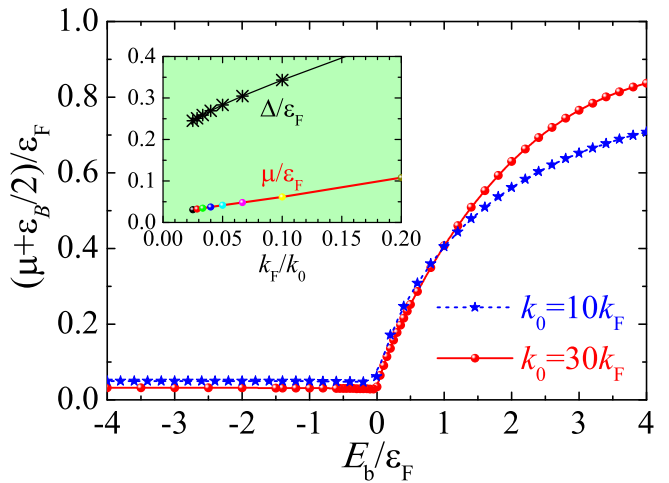}\caption{\label{fig8_k0dep} (color online). The chemical potential $\mu$
(with the two-body bound state contribution subtracted) as a function
of the scattering energy $E_{b}$, at two cut-off momenta $k_{0}=10k_{F}$
(dashed line with stars) and $k_{0}=30k_{F}$ (solid line with circles).
The inset shows the chemical potential $\mu$ and pairing gap $\Delta$
at the zero scattering energy $E_{b}=0$, as a function of the inverse
cut-off momentum $k_{0}^{-1}$. All the results are predicted by using
the GPF theory.}
\end{figure}

\subsection{The dependence on the cut-off momentum $k_{0}$}

We now turn to discuss the cut-off momentum dependence of our results.
In the main figure of Fig. \ref{fig8_k0dep}, we compare the chemical
potentials at the BEC-BCS evolution at two cut-off momenta, $k_{0}=10k_{F}$
(dashed line with stars) and $k_{0}=30k_{F}$ (solid line with circles).
A factor of three reduction in the cut-off momentum does not lead
to any changes at the qualitative level. In the inset, we highlight
the cut-off momentum dependence of the chemical potential and pairing
gap at the resonance. We do not find singular behaviors as we increase
the cut-off momentum and extend it towards infinity. Therefore, although
a cut-off momentum $k_{0}$ is necessary to make the $p$-wave interaction
renormalizable (for dimensions $d\geq2$), we may still have some
universal behaviors that are weakly (i.e., logarithmically) dependent
on $k_{0}^{-1}$.

\section{Conclusions and outlooks}

In conclusions, we have theoretically investigated the consequence
of quantum fluctuations in a resonantly interacting $p$-wave Fermi
superfluid in two dimensions at the BEC-BCS evolution, using the Gaussian
pair fluctuation theory. We have found that the zero-temperature equations
of state, the critical velocity for superfluidity, and the BKT critical
temperature are strongly renormalized by quantum fluctuations and
their non-analyticity at the topological phase transition is greatly
enhanced. Experimentally, this non-analyticity could be best probed
by measuring the pressure equation of state at zero temperature, which
shows an apparent kink near resonance, and the BKT critical temperature,
which exhibits a bump and then a cusp structure. Although the $p$-wave
Fermi superfluid seems to be delicate in superfluidity compared with
its $s$-wave counterpart due to a smaller critical velocity, it is
thermodynamically stable at all interaction strengths, in disagreement
with a previous theoretical study \cite{Jiang2018}, which takes into
account quantum fluctuations at the level of two-loop diagrams.

For $p$-wave interacting fermions in two dimensions, Nishida and
co-workers recently predicted the existence of a series of three-particle
bound states, the so-called super-Efimov states \cite{Nishida2013}.
The impact of these super-Efimov states to the many-body properties
(i.e., superfluidity) of the system remains to be understood. It will
be an interesting research topic to be explored in future studies. 
\begin{acknowledgments}
This research was supported by Australian Research Council's (ARC)
Programs FT130100815 and DP170104008 (HH), DE180100592 (JW), FT140100003
and DP180102018 (XJL), the National Key R\&D Program of China (Grant
No. 2018YFA0306503) (LH), and the National Natural Science Foundation
of China, Grant No. 11775123 (LH). 
\end{acknowledgments}

\appendix

\section{Two-particle scattering}

We use a separable interaction potential to characterize the chiral
$p$-wave interatomic interaction: 
\begin{eqnarray}
V_{{\bf kk}^{\prime}} & = & \lambda\Gamma\left({\bf k}\right)\Gamma^{*}\left({\bf k}^{\prime}\right),\\
\Gamma\left({\bf k}\right) & = & \frac{\left(k/k_{1}\right)}{\left[1+\left(k/k_{0}\right)^{2n}\right]^{3/2}}e^{i\varphi_{{\bf k}}},
\end{eqnarray}
where $k_{1}$ - set to be $k_{F}$ in numerical calculations - is
a characteristic momentum that makes $\Gamma\left({\bf k}\right)$
dimensionless, $k_{0}$ is a large-momentum cut-off, and $\varphi_{{\bf k}}$
is the polar angle of ${\bf k}$ in 2D momentum space. We use the
exponent $n$ to control the shape of the regularization function
$\Gamma\left({\bf k}\right)$. In the large-$n$ limit, effectively
we have a step function.

To obtain the two-body scattering amplitude, we consider the following
two-body $T$-matrix in vacuum, 
\begin{eqnarray}
T\left({\bf k},{\bf k}^{\prime};E_{+}\right) & = & t\left(E_{+}\right)\Gamma\left({\bf k}\right)\Gamma^{*}\left({\bf k}^{\prime}\right),\\
\frac{1}{t\left(E_{+}\right)} & = & \frac{1}{\lambda}+\sum_{{\bf k}^{\prime\prime}}\frac{\left|\Gamma\left({\bf k}^{\prime\prime}\right)\right|^{2}}{2\epsilon_{{\bf k}^{\prime\prime}}-E_{+}},
\end{eqnarray}
where $k^{\prime}=\left|{\bf k}^{\prime}\right|=k$ and $E_{+}=\hbar^{2}k^{2}/m+i0^{+}$.
The analytic form of the scattering amplitude or $t\left(E_{+}\right)$
in the low energy limit (i.e., $k\rightarrow0$) should be independent
on the detailed regularization function. Therefore, we may simply
use a step function (i.e., $n\rightarrow\infty$). By introducing
a new variable $x=(k^{\prime\prime})^{2}$, we find that, 
\begin{equation}
\frac{1}{t\left(E_{+}\right)}=\frac{1}{\lambda}+\frac{m}{4\pi\hbar^{2}k_{1}^{2}}\int\limits _{0}^{k_{0}^{2}}dx\frac{x}{x-\tilde{E}_{+}},
\end{equation}
where $\tilde{E}_{+}=k^{2}+i0^{+}$. This leads to ($\tilde{E}=k^{2}$),
\begin{widetext}
\begin{equation}
\frac{1}{t\left(E_{+}\right)}=\frac{1}{\lambda}+\frac{m}{4\pi\hbar^{2}k_{1}^{2}}\left[k_{0}^{2}+\tilde{E}\ln\left(\frac{k_{0}^{2}}{\tilde{E}}-1\right)+i\pi\tilde{E}\right].\label{eq:InverseTwoBodyTmatrix}
\end{equation}
By taking the low-energy limit $k\rightarrow0$, we arrive at 
\begin{equation}
\frac{1}{t\left(E_{+}\right)}=-\frac{m}{4\hbar^{2}k_{1}^{2}}\left[-\frac{1}{a_{p}}+\frac{2k^{2}}{\pi}\ln\left(R_{p}k\right)-ik^{2}\right],
\end{equation}
where $R_{p}\sim1/k_{0}$ is the effective range of the $p$-wave
interaction, the term $a_{p}^{-1}$ collects all the constants in
Eq. (\ref{eq:InverseTwoBodyTmatrix}) and physically we interpret
$a_{p}$ as the $p$-wave scattering area in two dimensions. It is
easy to see that, the full two-body $T$-matrix is ($k=k^{\prime}$),
\begin{equation}
T\left({\bf k},{\bf k}^{\prime};E_{+}\right)=e^{i\left(\varphi_{{\bf k}}-\varphi_{{\bf k}^{\prime}}\right)}\left(-\frac{4\hbar^{2}}{m}\right)\left[\frac{kk^{\prime}}{-a_{p}^{-1}+\left(2k^{2}/\pi\right)\ln\left(R_{p}k\right)-ik^{2}}\right].
\end{equation}
According to Levinsen, Cooper and Gurarie (see Appendix in Ref.\cite{Levinsen2008}),
we may define a two-dimensional $p$-wave scattering amplitude, 
\begin{equation}
f_{p}\left(k\right)=-\frac{m}{2\hbar^{2}\left(2\pi k\right)^{1/2}}T\left({\bf k},{\bf k};E_{+}\right)=\sqrt{\frac{2}{\pi k}}\frac{k^{2}}{-a_{p}^{-1}+\left(2k^{2}/\pi\right)\ln\left(R_{p}k\right)-ik^{2}}=\frac{1}{g_{p}(k)-i\left(\pi k/2\right)^{1/2}},
\end{equation}
\end{widetext}

where 
\begin{equation}
g_{p}(k)=\sqrt{\frac{\pi k}{2}}\frac{-a_{p}^{-1}+\left(2k^{2}/\pi\right)\ln\left(R_{p}k\right)}{k^{2}}
\end{equation}
is a real function of $k$. The $p$-wave scattering amplitude may
also be written in terms of the phase shift $\delta_{p}(k)$ \cite{Levinsen2008}:
\begin{equation}
f_{p}\left(k\right)=\frac{1}{i\sqrt{2\pi k}}\left(e^{2i\delta_{p}}-1\right)=\sqrt{\frac{2}{\pi k}}\frac{1}{\cot\delta_{p}-i},\label{eq:fsc}
\end{equation}
where the phase shift satisfies, 
\begin{equation}
k^{2}\cot\delta_{p}\left(k\right)=-\frac{1}{a_{p}}+\frac{2k^{2}}{\pi}\ln\left(R_{p}k\right)+\cdots.
\end{equation}
We note that, the relation between the scattering amplitude $f_{p}(k)$
and phase shift $\delta_{p}(k)$ defined in Eq. (\ref{eq:fsc}) is
slightly different from that derived by solving the two-body problem
(see Eq. (11) in Ref. \cite{Zhang2017})

\section{The structure of the functions $A$, $B$, $C$, $D$ and $F$}

Here we demonstrate that the functions $A$, $B$, $C$, $D$ and
$F$ do not depend on the direction of ${\bf q}$, and thus we may
simply set ${\bf q}=q{\bf e}_{x}$ in numerical calculations. Actually,
this is pretty clear for $A$, $B$, $C$ and $D$, since the factor
$\left|\Gamma({\bf k)}\right|$ does not depend on the polar angle
$\varphi_{{\bf k}}$. All the integral functions therefore depend
on the angle between ${\bf q}$ and ${\bf k}$ only, or more precisely
$\cos(\varphi_{{\bf k}}-\varphi_{{\bf q}})$. For the function $F$,
we now need to check explicitly that the factor 
\begin{equation}
P=\Gamma^{*}\left({\bf k}\right)\Gamma^{*}\left({\bf k}\right)\Gamma\left(\frac{\mathbf{q}}{2}+{\bf k}\right)\Gamma\left(\frac{\mathbf{q}}{2}-{\bf k}\right)
\end{equation}
also depends on $\varphi_{{\bf k}}-\varphi_{{\bf q}}$ only. We may
also explicitly show that $F$ is a real function. For this purpose,
we examine the following product,
\begin{widetext}
\begin{eqnarray}
P & = & \tilde{P}\left(k,q;\varphi_{{\bf k}}-\varphi_{{\bf q}}\right)e^{-i2\varphi_{{\bf k}}}\left[\left(\frac{q_{x}}{2}+k_{x}\right)+i\left(\frac{q_{y}}{2}+k_{y}\right)\right]\left[\left(\frac{q_{x}}{2}-k_{x}\right)+i\left(\frac{q_{y}}{2}-k_{y}\right)\right],\\
 & = & \tilde{P}\left(k,q;\varphi_{{\bf k}}-\varphi_{{\bf q}}\right)e^{-i2\varphi_{{\bf k}}}\left[\frac{\left(q_{x}^{2}-q_{y}^{2}\right)}{4}-\left(k_{x}^{2}-k_{y}^{2}\right)+i2\left(\frac{q_{x}q_{y}}{4}-k_{x}k_{y}\right)\right],\\
 & = & \tilde{P}\left(k,q;\varphi_{{\bf k}}-\varphi_{{\bf q}}\right)\left[\cos2\varphi_{{\bf k}}-i\sin2\varphi_{{\bf k}}\right]\left[\left(\frac{q^{2}}{4}\cos2\varphi_{{\bf q}}-k^{2}\cos2\varphi_{{\bf k}}\right)+i\left(\frac{q^{2}}{4}\sin2\varphi_{{\bf q}}-k^{2}\sin2\varphi_{{\bf k}}\right)\right]\\
 & = & \tilde{P}\left(k,q;\varphi_{{\bf k}}-\varphi_{{\bf q}}\right)\left\{ \left[\frac{q^{2}}{4}\cos\left(2\varphi_{{\bf k}}-2\varphi_{{\bf q}}\right)-k^{2}\right]-i\frac{q^{2}}{4}\sin\left(2\varphi_{{\bf k}}-2\varphi_{{\bf q}}\right)\right\} ,
\end{eqnarray}
\end{widetext}

where in the first line of the equation, we have singled out the chiral
$p_{x}+ip_{y}$ dependence of the regularization function $\Gamma$
and the function $\tilde{P}$ depends on $\varphi_{{\bf k}}-\varphi_{{\bf q}}$.
It is now clear that, in the calculations of $A$, $B$, $C$, $D$
and $F$, $\varphi_{{\bf q}}$ can be removed by re-defining the angle
$\varphi_{{\bf k}}$: $\varphi_{{\bf k}}-\varphi_{{\bf q}}\rightarrow\varphi$.
The imaginary part of $F$ is strictly zero since 
\begin{equation}
\int\limits _{0}^{2\pi}d\varphi h\left(\cos\varphi\right)\sin2\varphi=0
\end{equation}
for any function $h(x)$.

\section{Ginzburg-Landau free energy functional for the pair fluctuation field}

In the BEC limit, we may derive a Gross-Pitaevskii free energy of
composite bosons $\mathcal{S}[\phi(\mathbf{x},\tau)]$, which takes
the form, 
\begin{equation}
\mathcal{S}=\int dx\left[\phi^{*}\left(\frac{\partial}{\partial\tau}-\frac{\hbar^{2}}{2m_{B}}-\mu_{B}\right)\phi+\frac{g_{B}}{2}\left|\phi\right|^{4}\right],
\end{equation}
where $m_{B}=2m$ is the mass of composite bosons, $\mu_{B}$ is the
chemical potential and $g_{B}$ is the pair-pair interaction strength,
and we abbreviate $x\equiv(\mathbf{x},\tau)$. To this end, we first
consider the Ginzburg-Landau free energy functional for the pair fluctuation
field $\Delta(x)$: 
\begin{equation}
\mathcal{\tilde{S}}=\int dx\left[\Delta^{*}\left(a\frac{\partial}{\partial\tau}-b\frac{\hbar^{2}}{4m}-c\right)\Delta+\frac{d}{2}\left|\Delta\right|^{4}\right],
\end{equation}
where the $\phi$-field can be obtained by rescaling the pair fluctuation
field $\Delta$, i.e., $\sqrt{a}\Delta(x)\rightarrow\phi(x)$.

Following the seminal work by Sá de Melo, Randeria, and Engelbrecht
\cite{SadeMelo1993}, we determine the coefficients $a$, $b$, and
$c$ by evaluating the small frequency and momentum expansion of the
pair propagator $M_{0}(\mathbf{q},i\nu_{n})$ in the normal state,
which takes the form, 
\begin{equation}
M_{0}=-\frac{1}{2\lambda}+\frac{1}{2}\sum_{\mathbf{k}}\frac{\left|\Gamma(\mathbf{k})\right|^{2}}{i\nu_{n}+2\mu-2\epsilon_{\mathbf{k}}-\hbar^{2}\mathbf{q}^{2}/(4m)}.
\end{equation}
Using the fact that, 
\begin{equation}
M_{0}\left(\mathbf{q}\rightarrow0,i\nu_{n}\rightarrow0\right)\simeq-a\left(i\nu_{n}\right)+b\frac{\hbar^{2}\mathbf{q}^{2}}{4m}-c,
\end{equation}
we obtain 
\begin{equation}
a=b=\frac{1}{2}\sum_{\mathbf{k}}\frac{\left|\Gamma(\mathbf{k})\right|^{2}}{\left(2\epsilon_{\mathbf{k}}-2\mu\right)^{2}}
\end{equation}
and 
\begin{equation}
c=\frac{1}{2\lambda}+\frac{1}{2}\sum_{\mathbf{k}}\frac{\left|\Gamma(\mathbf{k})\right|^{2}}{2\epsilon_{\mathbf{k}}-2\mu}.
\end{equation}
In the BEC limit, we have $\mu_{B}=2\mu-E_{b}\rightarrow0^{+}$. By
replacing the bare interaction strength $\lambda$ with the scattering
energy $E_{b}$, it is easy to verify that, 
\begin{equation}
c\simeq\frac{\mu_{B}}{2}\sum_{\mathbf{k}}\frac{\left|\Gamma(\mathbf{k})\right|^{2}}{\left(2\epsilon_{\mathbf{k}}+\left|E_{b}\right|\right)^{2}}\simeq\mu_{B}a.
\end{equation}
The integral in $a$ can be worked out in the limit of an infinitely
large exponent $n\rightarrow\infty$, where $\Gamma(\mathbf{k})=[(k_{x}+ik_{y})/k_{1}]\Theta(k_{0}-k)$.
We find that, 
\begin{equation}
a=\frac{m^{2}}{8\pi\hbar^{4}k_{1}^{2}}\left(\ln\eta+\frac{1}{\eta}-1\right),
\end{equation}
where $\eta=\hbar^{2}k_{0}^{2}/(m\left|E_{b}\right|)+1$.

The coefficient $d$, on the other hand, may be calculated by Taylor
expanding the mean-field thermodynamic potential $\Omega_{\textrm{MF}}$
at small pairing gap $\Delta\sim0$, i.e., 
\begin{eqnarray*}
\Omega_{\textrm{MF}} & = & -c\Delta^{2}+\frac{d}{2}\Delta^{4}+\cdots.
\end{eqnarray*}
This leads to, 
\begin{equation}
c=-\frac{\partial\Omega_{\textrm{MF}}}{\partial\Delta^{2}}=\frac{1}{2\lambda}+\sum_{\mathbf{k}}\frac{\left|\Gamma(\mathbf{k})\right|^{2}}{2\epsilon_{\mathbf{k}}-2\mu}
\end{equation}
as anticipated, and 
\begin{equation}
d=\frac{\partial^{2}\Omega_{\textrm{MF}}}{\partial\left(\Delta^{2}\right)^{2}}=\sum_{\mathbf{k}}\frac{\left|\Gamma(\mathbf{k})\right|^{4}}{\left(2\epsilon_{\mathbf{k}}-2\mu\right)^{3}}.
\end{equation}
By replacing $-2\mu$ with $\left|E_{b}\right|$ in the equation for
$d$, and performing the integration, we obtain, 
\begin{eqnarray}
d & = & \frac{m^{3}}{4\pi\hbar^{6}k_{1}^{4}}\left(\ln\eta+\frac{2}{\eta}-\frac{1}{2\eta^{2}}-\frac{3}{2}\right).
\end{eqnarray}
The rescaling of the pair fluctuation field, $\sqrt{a}\Delta(x)\rightarrow\phi(x)$,
leads to the desired expression for the pair-pair interaction strength,
\begin{equation}
g_{B}=\frac{d}{a^{2}}=\frac{16\pi\hbar^{2}}{m}\frac{\left[\ln\eta+2\eta^{-1}-\eta^{-2}/2-3/2\right]}{\left(\ln\eta+\eta^{-1}-1\right)^{2}},
\end{equation}
which is Eq. (\ref{eq:gBMF}) in the main text.


\begin{thebibliography}{10}
\bibitem{TopoSuperfluidNote}In order to have a smoking gun proof
of the non-trivial topological properties of the Fermi superfluid,
it is necessary to confirm the existence of the chiral Majorana mode
at the edge or the non-abelian Majorana fermions at the vortex core.
\end{thebibliography}
\end{document}